\documentclass{emulateapj}
\usepackage{graphicx}
\usepackage[usenames]{color}
\usepackage{natbib}
\bibpunct{(}{)}{:}{a}{}{,}

\shorttitle{Red giant in an eclipsing binary system from \textit{Kepler}}
\shortauthors{S. Hekker et al.}

\begin{document}


\title{Discovery of a red giant with solar-like oscillations in an eclipsing
binary system from \textit{Kepler} space-based photometry}

\author{S. Hekker\altaffilmark{1}, 
J. Debosscher\altaffilmark{2},
D. Huber\altaffilmark{3}, 
M.~G. Hidas\altaffilmark{3,4,5}, 
J. De Ridder\altaffilmark{2}, 
C. Aerts\altaffilmark{2,6}, 
D. Stello\altaffilmark{3},
T.R. Bedding\altaffilmark{3}, 
R.~L. Gilliland\altaffilmark{7},
J. Christensen-Dalsgaard\altaffilmark{8}, 
T.~M. Brown\altaffilmark{4},
H. Kjeldsen\altaffilmark{8}, 
W.~J. Borucki\altaffilmark{9},
D. Koch\altaffilmark{9},
J.~M. Jenkins\altaffilmark{10},
H. Van Winckel\altaffilmark{2},
P.~G. Beck\altaffilmark{2}, 
J. Blomme\altaffilmark{2},
J. Southworth\altaffilmark{11},
A. Pigulski\altaffilmark{12}, 
W.~J. Chaplin\altaffilmark{1}, 
Y.~P. Elsworth\altaffilmark{1},
I.~R. Stevens\altaffilmark{1}, 
S. Dreizler\altaffilmark{13},
D.~W. Kurtz\altaffilmark{14},
C. Maceroni\altaffilmark{15}, 
D. Cardini\altaffilmark{16},
A. Derekas\altaffilmark{3,17},
M.~D. Suran\altaffilmark{18} } 
\email{saskia@bison.ph.bham.ac.uk}

\altaffiltext{1}{School of Physics and Astronomy, University of Birmingham,
Edgbaston B15 2TT, United Kingdom}
\altaffiltext{2}{Instituut voor Sterrenkunde, Katholieke Universiteit Leuven,
Celestijnenlaan 200D, B-3001 Leuven, Belgium}
\altaffiltext{3}{Sydney Institute for Astronomy (SIfA), School of Physics,
University of Sydney, NSW 2006, Australia}
\altaffiltext{4}{Las Cumbres Observatory Global Telescope, Goleta, CA 93117,
USA} 
\altaffiltext{5}{Department of Physics, University of California, Santa
Barbara, CA 93106, USA}
\altaffiltext{6}{IMAPP, Department of Astrophysics, Radbout University Nijmegen,
PO Box 9010, 6500 GL Nijmegen, the Netherlands}
\altaffiltext{7}{Space Telescope Science Institute, 3700 San Martin Drive,
Baltimore, MD 21218, USA}
\altaffiltext{8}{Department of Physics and Astronomy, Aarhus University, DK-8000
Aarhus C, Denmark}
\altaffiltext{9}{NASA Ames Research Center, MS 244-30, Moffet Field, CA 94035, USA}
\altaffiltext{10}{SETI Institute/NASA Ames Research Center, MS 244-30, Moffet Field, CA 94035, USA}
\altaffiltext{11}{Astrophysics Group, Keele University Newcastle-under-Lyme, ST5
5BG, UK}
\altaffiltext{12}{Instytut Astronomiczny Uniwersytetu Wroclawskiego, Kopernika
11, 51-622 Wroclaw, Poland}
\altaffiltext{13}{Georg-August Universit\"{a}t, Institut f\"{u}r Astrophysik,
Friedrich-Hund-Platz 1, D-37077 G\"{o}ttingen}
\altaffiltext{14}{Jeremiah Horrocks Institute of Astrophysics, University of
Central Lancashire, PR1 2HE, UK}
\altaffiltext{15}{INAF-Osservatorio Astronomica di Roma, via Frascati 3, I-00040
Monteporzio C. (RM), Italy}
\altaffiltext{16}{IASF-Roma, INAF, V. del Fosso del Cavaliere 100, 00133 Roma,
Italy}
\altaffiltext{17}{Konkoly Observatory, Hungarian Academy of Sciences, H-1525 Budapest, P.O.
Box 67, Hungary}
\altaffiltext{18}{Astronomical Institute of the Romanian Academy, Str. Cutitul
de Argint 5, RO 40557, Bucharest, RO}

\begin{abstract}
Oscillating stars in binary systems are among the most interesting stellar laboratories, as these can provide information on the stellar parameters and stellar internal structures. Here we
present a red giant with solar-like oscillations in an eclipsing binary observed
with the NASA \textit{Kepler} satellite. We compute stellar parameters of the red giant
from spectra and the asteroseismic mass and radius from the
oscillations. Although only one eclipse has been observed so far, we can already
determine that the secondary is a main-sequence F star in an eccentric
orbit with a semi-major axis larger than 0.5\,AU and orbital period longer than
75\,days.
\end{abstract}


\section{Introduction}

The prospects for asteroseismology of red-giant stars have increased
significantly with the launch of \textit{Kepler} \citep{borucki2009} in March 2009 and
CoRoT \citep{baglin2006} in December 2006. These satellites provide
uninterrupted long time series of high-precision photometry for a large sample
of stars. Among these stars are many red giants, allowing for both statistical
analyses as well as detailed individual studies. CoRoT data have already
revealed the presence of nonradial oscillation modes in red giants
\citep{deridder2009} and have given rise to a population synthesis study
\citep{miglio2009}. The currently available \textit{Kepler} data reveal clear solar-like
high radial overtone p-mode oscillations in a large sample of red giants,
extending in luminosity from the red clump to the bottom of the giant branch
\citep{bedding2010}.

For in-depth studies, including detailed stellar modeling, accurate stellar
parameters are needed. Red giants of different masses occupy a narrow region in
the H-R diagram, which results in relatively large errors when estimating the
mass from measured classical parameters, such as effective temperature, surface
gravity and luminosity, and from evolutionary tracks. A better estimate can be
obtained from the characteristic frequencies of solar-like oscillations in these
stars and their frequency separations, either through modeling or through
scaling relations \citep{kjeldsen1995}. Here we use a modeling approach and
scaling relations to compute an asteroseismic mass and radius.

Eclipsing binary systems provide the primary method for measuring masses and
radii of stars. However, the probability of observing an eclipsing system is
low, because of the geometrical restriction that the system has to be viewed
close to the orbital plane. We report here the discovery of a pulsating red
giant in an eclipsing binary system observed by \textit{Kepler}. This star was found
using automated variability classification methods to the targets of the
asteroseismology program of \textit{Kepler} \citep{debosscher2009, blomme2010}.

\section{Observations}

\begin{figure}
\begin{minipage}{\linewidth}
\centering
\includegraphics[width=\linewidth]{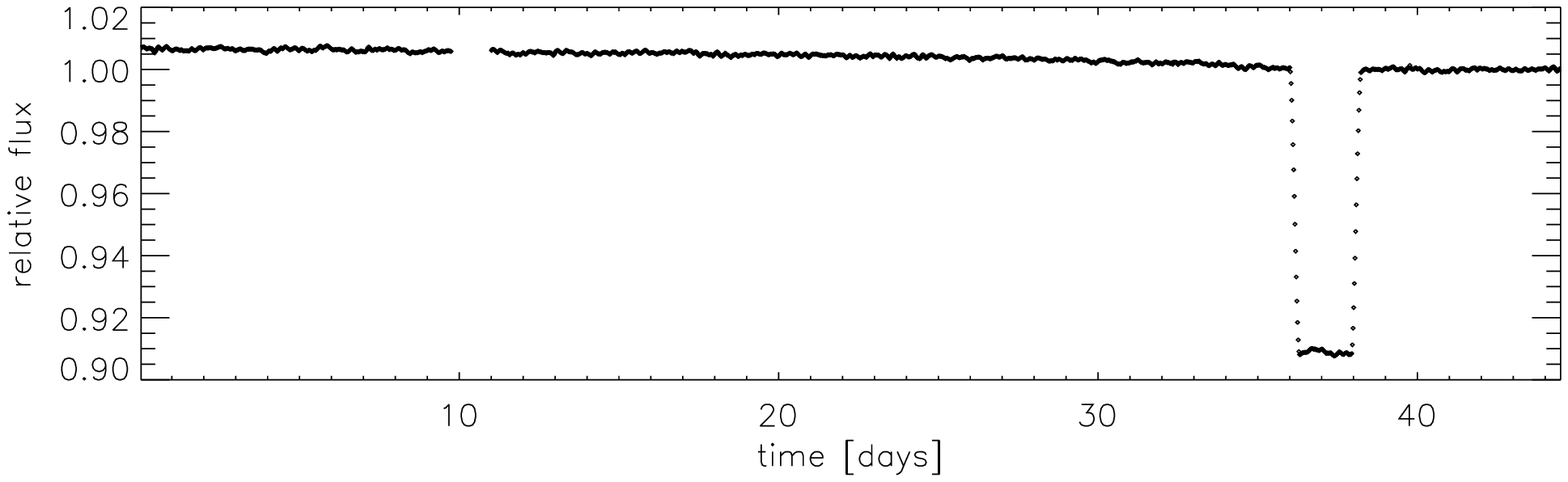}
\end{minipage}
\begin{minipage}{\linewidth}
\centering
\includegraphics[width=\linewidth]{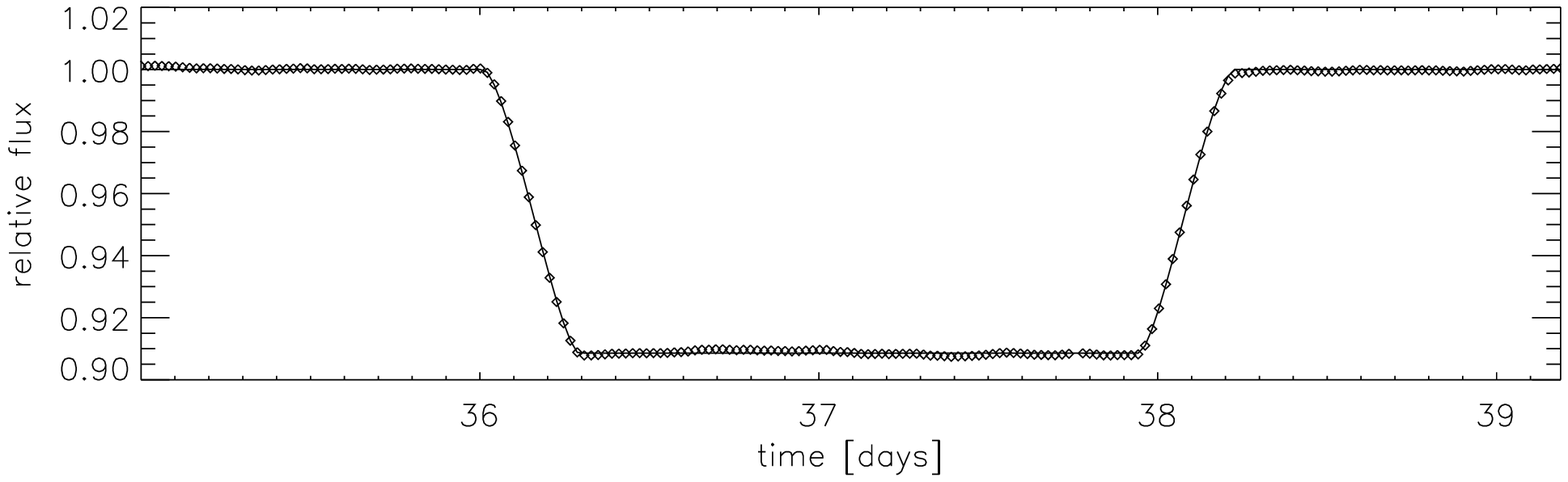}
\end{minipage}
\begin{minipage}{\linewidth}
\centering
\includegraphics[width=\linewidth]{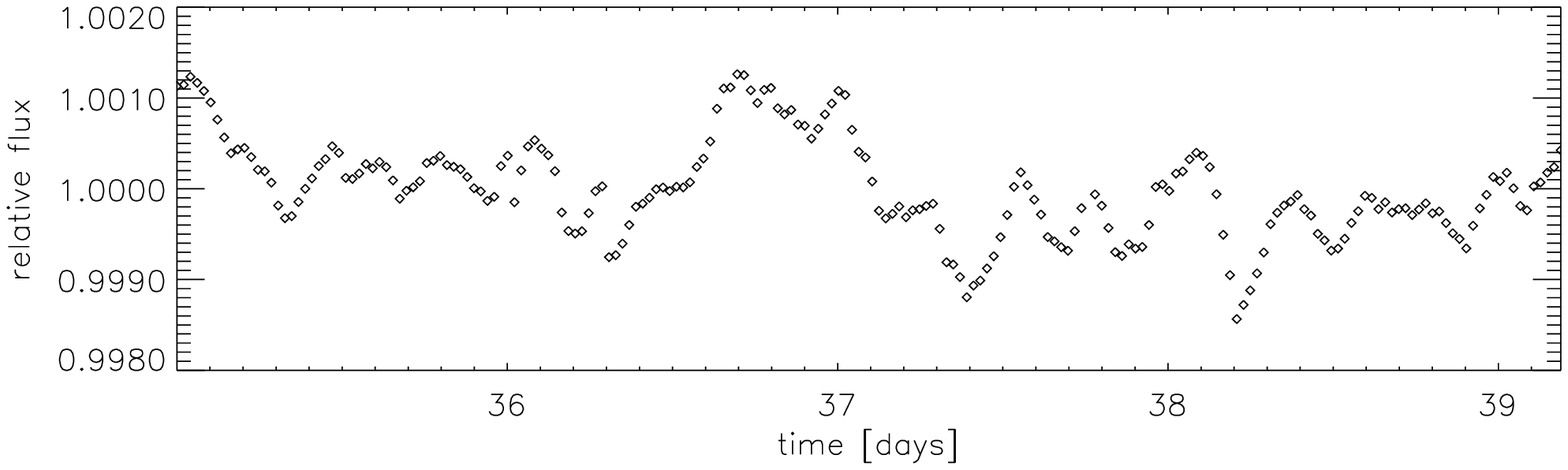}
\end{minipage}
\caption{Top: time series data from \textit{Kepler} of KIC8410637. Middle: a section of
the light curve centered on the eclipse with the best-fit model (solid line). Bottom: the residuals after correcting for the eclipse model, clearly showing the solar-like oscillations.}
\label{obs}
\end{figure}

Photometric data for KIC8410637\footnote{KIC = \textit{Kepler Input
Catalog}, \citet{brown2005}.} (TYC 3130-2385) have been taken with \textit{Kepler} during
the commissioning run (Q0) from 2009 May 2 through 11, and the following first
science run (Q1) of 33.5\,d. We present results based on these combined time
series of about 40\,d. This star has a magnitude of 10.77 mag in the \textit{Kepler}
bandpass and was selected by the \textit{Kepler} Asteroseismic Science Consortium \citep[KASC,][]{gilliland2009} for long-cadence observations. The time series data have a 29.4-min cadence and
show both solar-like oscillations in the frequency regime expected for a red
giant, and an eclipse (see Fig.~\ref{obs}).

This star has also been part of the TrES-Lyr1 ground-based
survey\footnote{\tt http://nsted.ipac.caltech.edu/NStED/docs/holdings.html}
\citep{odonovan2006} from which time series photometry with a time span of 75\,d
are at our disposal. In these data no eclipse is visible, while the primary and
possibly the secondary eclipse (depth predicted in Section 6) would have been
detected, had they occurred during the time span of the observations. The fact
that the primary is not seen in the TrES data implies that the period is longer
than the duration of the TrES observations.

This star was also observed photometrically by the SuperWASP cameras
\citep{pollacco2006} for 100 days in 2007 and for 90 days in 2008. There is a
suspicion of a secondary eclipse, but no primary eclipse has been observed during
these periods, which have some gaps due to weather conditions. Observations for
KIC8410637 are also available from the All Sky Automated Survey (ASAS)
\citep{pojmanski1997}, in which there are three points that have lower flux than
the majority. These points might be due to the eclipse, which would imply an
orbital period of about 380\,d or a submultiple thereof. Unfortunately, there
are too few points, and their uncertainty is too large to draw firm conclusions.

Additionally, we checked whether we could put more constraints on the period from the time stamps of the data we have at our disposal. We determined which periods would have led us to miss primary eclipses because of a lack of observations. We would have missed all eclipses for an orbital period of about 135, 220, 253, 261, 270 and 295 days. All other possible periods are longer than the 380 days, mentioned above.

\begin{table}
\begin{minipage}{\linewidth}
\caption{Stellar parameters of KIC8410637.}
\label{param}
\centering
\begin{tabular}{lccc}
\hline\hline
 & KIC & \citet{ofek2008} & spectroscopy \\
 \hline
$T_{\rm  eff}$ [K] & 4680 $\pm$ 150 & 4650 & 4650 $\pm$ 80\\
$\log$(g) (c.g.s.) & 2.8 $\pm$ 0.3 & & 2.70 $\pm$ 0.15\\
$\rm [Fe/H]$ [dex] & $-$0.3 $\pm$ 0.3 & $-$0.38 & 0.0 $\pm$ 0.1\\
$v\sin i$ [km\,s$^{-1}$] & & & 5.0 $\pm$ 0.7 \\
$\xi_{\rm micro}$ [km\,s$^{-1}$] & & & 1.3 $\pm$ 0.1\\
\hline
\hline
\end{tabular}
\end{minipage}
\end{table}

\section{Stellar parameters}

For KIC8410637 the \textit{Kepler Input Catalogue} \citep{brown2005} provides a set of
stellar parameters, as listed in Table~\ref{param}. Tycho-2 \citep{hog2000} and
2MASS \citep{skrutskie2006} photometry are also available. \citet{ofek2008} fitted these colors with synthetic photometry of stellar
templates \citep{pickles1998}. Using two methods, \citet{ofek2008} classified
the star as a metal-weak K2III star with an effective temperature of 4650\,K or
a K4V star with a temperature of 4350\,K.

We also obtained three spectra with the HERMES spectrograph \citep{raskin2008}
mounted on the Mercator Telescope, La Palma, Spain: two on 2009 October 13 and
one on 2009 October 15. These spectra have a resolution of about 85\,000 and a
wavelength range of $4000 - 9000$\,\AA. We analyzed the average spectrum, which
has a typical signal-to-noise ratio of $\sim$ 90 at 6000 \AA, with methods
described by \citet{hekker2007} and list the results in the last column of
Table~\ref{param}. This method does not directly provide uncertainties on the
individual parameters. Here, we adopted the total of the difference and scatter
between the values of \citet{hekker2007} and the values of \citet{luck2007}
\citep[see][for more details]{hekker2007}. For $v\sin i$, we computed the
uncertainty from a 10\% uncertainty in the total FWHM of the lines, which is
combined with the $\sigma$ on the macro turbulence. These results are in
agreement with those in the KIC and the literature. They confirm the red-giant
nature of the primary star.

We investigated the spectrum for signatures of the secondary star, in the form
of additional absorption lines and cross-correlation profiles, but no clear
signatures could be detected. The three spectra obtained over a time span of 3
days are not sufficient to see radial velocity variations due to the binary
orbit.

\section{Asteroseismic analysis}

\begin{figure}
\begin{minipage}{\linewidth}
\centering
\includegraphics[width=\linewidth]{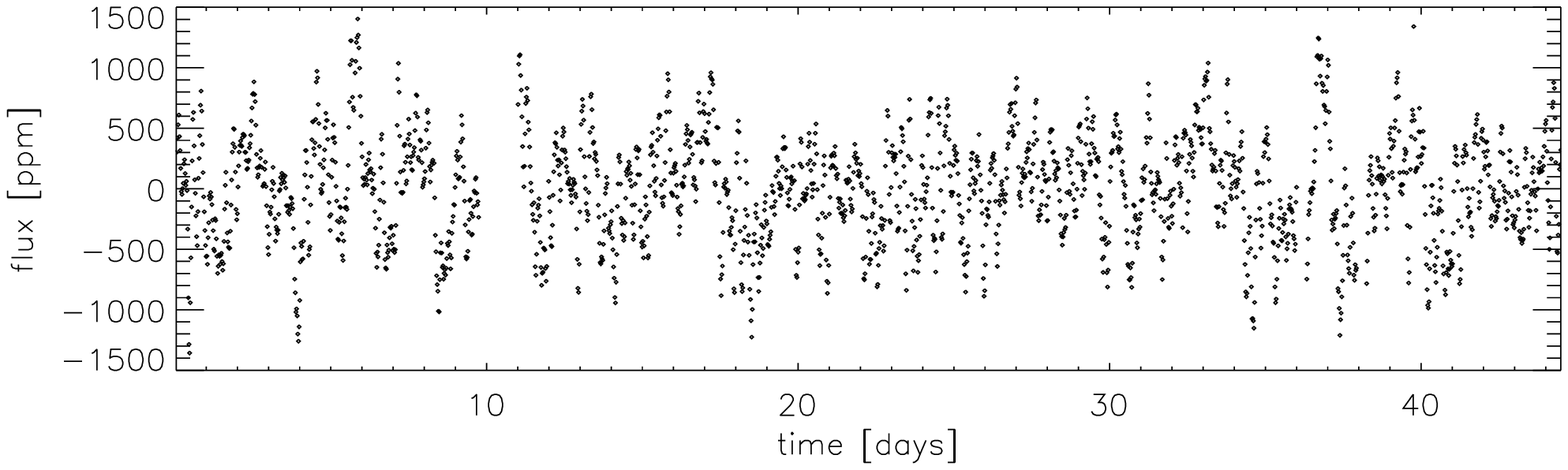}
\end{minipage}
\begin{minipage}{\linewidth}
\centering
\includegraphics[width=\linewidth]{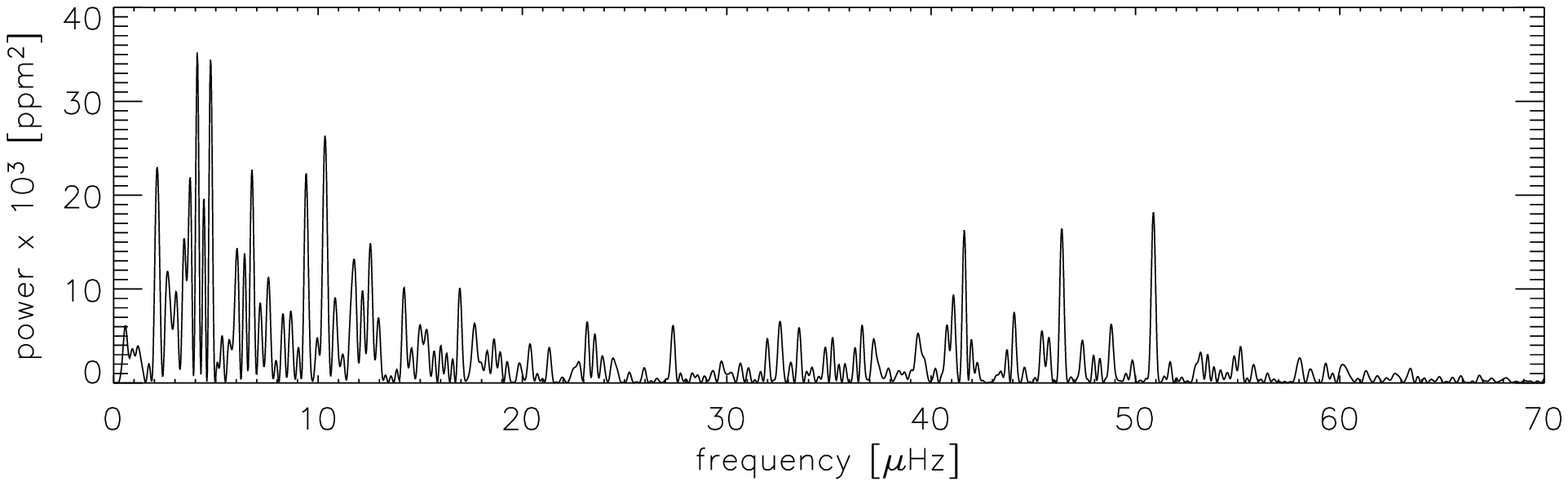}
\end{minipage}
\begin{minipage}{\linewidth}
\centering
\includegraphics[width=\linewidth]{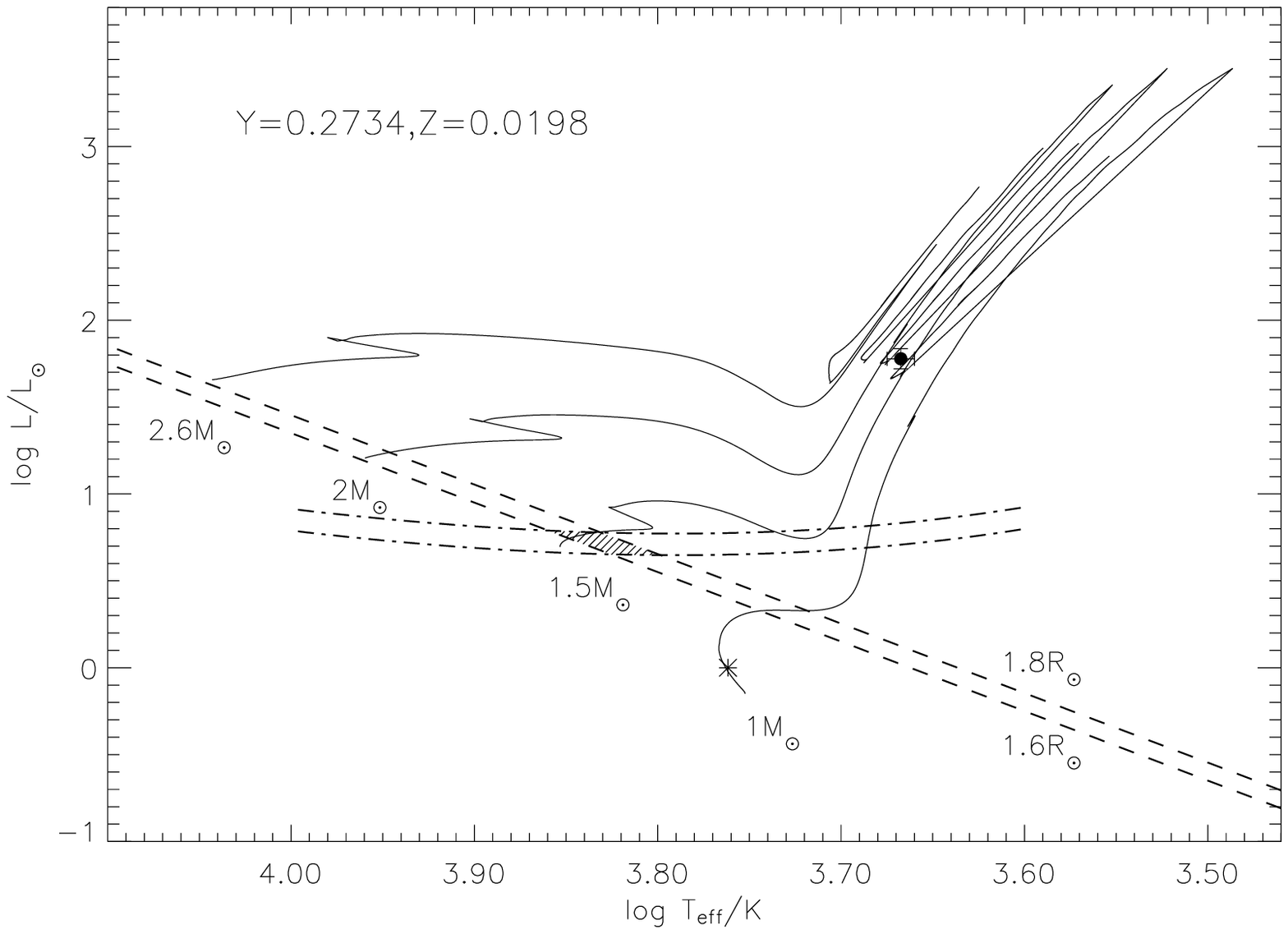}
\end{minipage}
\caption{Top: flux as a function of time, after removing the binary signature as
described in the text. Centre: the power spectrum of the red giant
oscillations. The power excess due to solar-like oscillations is present in the
range $30-60$\,$\mu$Hz. The excess below 20\,$\mu$Hz might be due to granulation in the primary's atmosphere. Bottom: H-R diagram with BASTI \citep{pietrinferni2004} evolutionary tracks
for different masses and solar metallicity. The filled circle shows the position
of the red giant.  The shaded area shows the secondary position
constrained by two values derived from the light-curve fit: the radius ratio (dashed lines) and the bolometric luminosity ratio (dashed-dotted lines; taking into account the \textit{Kepler} bandpass). The asterisk indicates the position of the Sun.}
\label{fluxosc}
\end{figure}

Solar-like oscillations are visible both during and outside the eclipse
(Fig.~\ref{obs}). To analyze these we adopted two approaches. In the first
approach, we corrected the light curve including the eclipse by fitting and
subtracting trends for different parts of the time series using a linear
polynomial for the commissioning data, a second-order polynomial for the part of
the Q1 data before the eclipse, the mean value for the data in full eclipse, and
a linear polynomial for the data obtained after the eclipse. We omitted the
observations during ingress and egress. The resulting time series is shown in
the top panel of Fig.~\ref{fluxosc}. In the second approach, we omitted the complete eclipse from the time
series. 

The Fourier spectrum shows a clear power excess (see the second panel of
Fig.~\ref{fluxosc}). The power excess is present between $30-60$\,$\mu$Hz, which
is the dominant range of $\nu_{\rm max}$ (frequency of maximum oscillation
power) found for a large sample of red giants observed by CoRoT
\citep{hekker2009a}. This reflects the high population density of red-clump
stars, compared to stars on the ascending branch \citep{miglio2009}. No power
excess is present at frequencies between 70 $\mu$Hz and the Nyquist frequency of
$\sim$ 280\,$\mu$Hz. We analyzed the power spectrum obtained from the corrected
time series with two methods of the Octave pipeline (\citealt{hekker2009}:
Octave1 and Octave2) and the power spectrum of the light curve where the eclipse
had been omitted with the automated pipeline developed in Sydney
\citep{huber2009}.

Octave1 computed $\nu_{\rm max}$ and $A_{\ell=0}$ (maximum mode amplitude) from
a binned power spectrum, and $\Delta \nu$ (large frequency separation) using the
power spectrum of the power spectrum. Octave2 computed $\nu_{\rm max}$ and
$A_{\ell=0}$ from a Gaussian fitting to the oscillation envelope, and $\Delta
\nu$ from an autocorrelation of the oscillation frequencies. The latter have
been determined with a Bayesian method, where oscillation frequencies are those
with a posterior probability of less than 0.1\% of being noise. The Sydney
pipeline produced $\nu_{\rm max}$ from a smoothed power spectrum. The current
data do not yet allow us to measure the life times of the solar-like
oscillations. For more details on the pipelines we refer to \citet{hekker2009}
and \citet{huber2009}.

The resulting parameters are listed in Table~\ref{oscparam}. Note that
$A_{\ell=0}$ is computed for a power spectrum with mean squared scaling, which
is a factor $\sqrt2$ different with respect to the scaling relations of
\citet{kjeldsen1995}.  With the effective temperature and the scaling laws from
\citet{kjeldsen1995} we computed the radius, mass and luminosity of the primary
(see Table~\ref{oscparam}).

We also used the RADIUS pipeline described by \citet{stello2009} to compute the
stellar radius. The input parameters for this pipeline are the large separation,
$T_{\rm eff}$ and metallicity, for which we used the spectroscopic values listed
in Table~\ref{param}. The results from this analysis are listed in
Table~\ref{oscparam} and lead to a value for $\log$(g) of 2.6 $\pm$ 0.1, which
is in agreement with our spectroscopically derived value.

\begin{table*}
\begin{minipage}{\linewidth}
\caption{Oscillation parameters of KIC8410637 computed with different methods. }
\label{oscparam}
\centering
\begin{tabular}{lcccc}
\hline\hline
 & Octave1 & Octave2 & Sydney & RADIUS \\
 &     ---        & measured &    ---     & model\\
 \hline
$\nu_{\rm max}$ [$\mu$Hz] & 45.0 $\pm$ 4.8 & 45.2 $\pm$ 1.3 & 44.2 $\pm$ 0.5 &\\
$\Delta \nu$ [$\mu$Hz] & 4.64 $\pm$ 0.23 & 4.5 $\pm$ 0.1 & 4.51 $\pm$ 0.37&\\
$\delta \nu_{02}$ [$\mu$Hz] & 0.6 $\pm$ 0.1 & & 0.69 $\pm$ 0.09&\\
$A_{\ell=0}$ [ppm] & 57.6 $\pm$ 7.3 & 61.8 $\pm$ 6.2 & 53.4 $\pm$ 1.9 & \\
&     ---        & derived &    ---     &\\
$R$ [R$_{\odot}$] & 11.2 $\pm$ 1.6 &11.8 $\pm$ 0.7 & 11.4 $\pm$ 1.9 & 13.1 $\pm$ 1.7\\
$L$ [L$_{\odot}$] &  53 $\pm$ 16 & 58 $\pm$ 8 & 55 $\pm$ 19 & 72 $\pm$ 19\\
$M$ [M$_{\odot}$] & 1.7 $\pm$ 0.7 & 1.8 $\pm$ 0.3 & 1.7 $\pm$ 0.9 & \\
\hline
\hline
\end{tabular}
\end{minipage}
\end{table*}

The values computed with different methods and approaches are in good agreement.
The weighted mean values of the results from the different methods are used
throughout the rest of the paper, i.e., 
$R_{\rm R}$~=~11.8~$\pm$~0.6~R$_{\odot}$,
$L_{\rm R}$~=~58~$\pm$~6~L$_{\odot}$, $M_{\rm R}$~=~1.7~$\pm$~0.3~M$_{\odot}$
(suffix `R' refers to the red giant) together with the spectroscopically
determined effective temperature. We point out  that the estimates of these fundamental parameters rely on the input physics of stellar models, through the interpretation of observables (e.g. $\nu_{\rm max}$ and $A_{l=0}$) in terms of the properties of the convection and thermodynamics, which we assumed to be known and error-free. This implies that there may be unknown systematic uncertainties. Nevertheless, the radius and mass have values in good agreement with the expected range for red-clump stars. For the radius there is also agreement between the result from scaling relations and from models (RADIUS).

\section{Binary analysis}

The very flat bottom of the eclipse, except for the red-giant oscillations,
seems to be indicative of a smaller object occulted by a considerably larger
object. This is underlined by the fact that for an annular eclipse,
i.e., transit, limb darkening effects play a more prominent role and would have
caused a more gradually changing shape of the 
eclipse.  Because we have only one eclipse, a full
analysis of the binary is not yet possible, but we are able to put several
constraints on the nature of the secondary and on the orbit.

From the \textit{Kepler} data, we directly measured the depth of the eclipse to be about
9\%. This depth is determined by the luminosity ratio of the two stars, i.e.,
$l = L_2/L_{\rm R}$ = 0.099 $\pm$ 0.005. The uncertainty is estimated taking the
shape of the light curve outside the eclipse into account. In addition to the
depth, we measured the time during which the secondary is fully obscured: $1.64
\pm 0.01$\,d. The ingress and egress are symmetric and each lasts for $0.276 \pm
0.007$\,d. The symmetry of the ingress and egress indicates that the relative acceleration of the binary components between ingress and egress is equal to zero, i.e. it is justified to assume the sky-projected relative velocity of the two stars to be constant during the eclipse.

The eclipse time depends on the radii of both stars and the geometry of the
system. Knowing that the eclipse lasts only during $\sim$~2\% or less of the
total orbital period, i.e., only a very small segment of the orbit is
covered during the eclipse, and that the system is viewed close to the orbital plane, implies that the time of total eclipse ($\tau_{\rm total}$) can be expressed as:
\begin{equation}
\tau_{\rm total}=\frac{P(R_{\rm R} \cos\delta - R_2)}{4aE(\epsilon)},
\label{tautot}
\end{equation}
with $P$ the orbital period, $a$ the semi-major axis, $E$ the complete elliptic integral, $\epsilon$ the eccentricity and $\delta$ the latitude of
the eclipse on the stellar disc and $R_{\rm R}$ and $R_2$ the radius of the red
giant and the secondary, respectively.  The total time of ingress, total eclipse
and egress ($\tau_{\rm in+total+eg}$) from the same geometric considerations can be
expressed as:
\begin{equation}
\tau_{\rm in+total+eg}=\frac{P(R_{\rm R} \cos\delta + R_2)}{4aE(\epsilon)}.
\label{tautotal}
\end{equation}

Dividing Eq.~(\ref{tautot}) by Eq.~(\ref{tautotal}) provides an estimate of
$R_2$ as a function of $\cos \delta$:
\begin{equation}
\frac{\tau_{\rm total}}{\tau_{\rm in+total+eg}} \equiv x = \frac{R_{\rm R}
\cos\delta - R_2}{R_{\rm R}\cos\delta + R_2} = \frac{\cos \delta - r}{\cos
\delta +r},
\label{tauovertau}
\end{equation}
with $r$ = $R_2/R_{\rm R}$. In this way we found that 
\begin{equation}
\frac{R_2}{ R_{\rm R}} = \cos\delta \frac{1-x}{1+x},
\end{equation}
where $x = 0.75 \pm 0.03$. Using $\delta$ = 0 (central eclipse) and the value of the radius of the red-giant star derived in the previous section, we then find an upper limit of $R_2 \le 1.7 \pm
0.1$\,R$_{\odot}$.

To confirm this geometric analysis we computed a basic model of the eclipse signal. We neglected limb darkening and computed the fractional overlap
between the two discs, taking into account the relative luminosity of the star
being eclipsed. As explained above, we assumed the sky-projected relative velocity ($v$) of the two stars to be constant. This model has been fitted to the portion of the light
curve near the eclipse using a Markov-Chain Monte Carlo approach \citep[see
e.g.][and references therein]{torres2008}, which yields the joint posterior
probability density of all the parameters, given the data.

The physical parameters in our model are $r$, $v$, $l$, and $b=\sin\delta$ (the
impact parameter). In order to avoid strong correlations among these parameters
(in particular between $r$, $v$, and $b$), when performing the fit we replace
$v$ with $v^2/(1-b^2)$ and $r$ with $r/v^2$.  The values of the physical
parameters resulting from this fit are $b=0.0^{+0.15}$,
$r=0.1435_{-0.0027}^{+0.0008}$, $v=1.0338_{-0.011}^{+0.0005} R_{\rm R}$/day, and
$l=0.10060_{-0.00006}^{+0.00006}$, consistent with the values obtained from the geometric analysis.

To test our assumption that $v$ is constant, we repeated the fit with
allowing $v$ to change linearly with time. The result was a deceleration
of $0.015_{-0.004}^{+0.005} R_{\rm R}/\rm day^2$ (i.e. $\sim 3$\% slower at
egress than at ingress). This may be a real asymmetry in the eclipse,
implying an eccentric orbit. However, it could also be due to the
pulsation signal altering the apparent shapes of the ingress and
egress.

We converted the fitted value of $l$ into a bolometric luminosity ratio by folding blackbody curves with the \textit{Kepler} spectral response curve. This is a function of the secondary's temperature, as shown in Fig.~\ref{fluxosc} (dot-dashed lines). With the constraint from the radius ratio and 
\begin{equation}
\frac{L_2}{L_{\rm R}} = \frac{R_2^2T_{\rm eff,2}^4}{R_{\rm R}^2T_{\rm eff,R}^4},
\label{radius}
\end{equation}
we find $L_2$ = 5.2 $\pm$ 0.7 L$_{\odot}$. 
The secondary is thus probably an F main-sequence star. Using the
mass-luminosity relation for stars with masses between 0.5 and 2.0\,M$_{\odot}$,
i.e., $(L/L_{\odot})=(M/M_{\odot})^{4.5}$, we found a
mass for the secondary $M_2$ of $1.44 \pm 0.05$\,M$_{\odot}$, where we did not
take the uncertainty of the mass-luminosity relation into account. We also
computed the surface gravity for the secondary
and found $\log$(g$_2$) = 4.2 $\pm$ 0.1.


The binary also has to obey Kepler's third law:
\begin{equation}
M_{\rm R}+M_2=\frac{4 \pi^2 a^3}{GP^2},
\label{kepler}
\end{equation}
in which we still have $a$ as unknown, and we take a minimum value of 75\,d for
$P$. With this value we computed a minimum semi-major axis for the system of
0.5\,AU.

On the other hand, if we assume a circular orbit, we can compute the
period. From the eclipse fit we know the orbital velocity and combining this
with Kepler's third law, we get:
\begin{equation}
P =  \frac{2 \pi G (M_{\rm R} +M_2)}{v^3}.
\label{kepler2}
\end{equation}
Using the masses of both stars, we obtain $P$ = 32 $\pm$ 6 days. This is considerably shorter than the minimum orbital period we inferred from currently available observations, which implies
an eccentric orbit.

\section{Discussion}

From observations obtained with \textit{Kepler}, we found a pulsating red giant in an
eclipsing binary. Once a full orbit has been observed, the orbital parameters
will provide an independent measure of the stellar parameters, such as mass and
radius, with respect to the asteroseismic values. These additional constraints
make this a very interesting case and the star is therefore being followed up by
the \textit{Kepler Mission} as well as with ground-based spectroscopy.

Stellar parameters of the primary red giant have been computed from a
spectroscopic analysis and from the solar-like oscillations. We did not find
evidence of the secondary component in the spectra. The currently available
spectra are not yet sufficient to put constraints on the orbit.

From the single eclipse observed by \textit{Kepler} and additional independent
observations from TrES, SuperWASP and ASAS, we inferred constraints on the orbit
and secondary star. The annular eclipse, i.e., the secondary passing in front of
the red giant, would cause an eclipse with a depth of $\sim$ 1.7\%, assuming a
circular orbit and without taking limb-darkening into account. Such a dip would
be observable, but has not been detected by \textit{Kepler}, TrES or ASAS. Only the
SuperWASP data show a feature that might be due to an annular eclipse. The
SuperWASP and ASAS data have gaps due to weather conditions. Although there are
also three gaps (4.7, 2.9 and 1.7\,d) in the TrES time series data with a
duration longer than the duration of the eclipse, we expect the chance that an
eclipse falls exactly during one of these gaps to be low. Therefore, we infer
that the orbit is longer than the time span of the TrES data, i.e., at least
75\,d. Also, an initial inspection of the raw \textit{Kepler} data of Q2 (additional 3 months of data following the Q1 phase) does not seem to reveal another eclipse. In the case that we missed an eclipse due to a lack of observations, the minimum orbital period would be 135 days.

\begin{table}
\begin{minipage}{\linewidth}
\caption{parameter overview of eclipsing binary KIC8410637. }
\label{allres}
\centering
\begin{tabular}{lcc}
\hline\hline
 & primary & secondary\\
\hline
$R$ [R$_{\odot}$] & 11.8 $\pm$ 0.6 & 1.7 $\pm$ 0.1\\
$L$ [L$_{\odot}$] & 58 $\pm$ 6  & 5.2 $\pm$ 0.7\\
$M$ [M$_{\odot}$] & 1.7 $\pm$ 0.3 & 1.44 $\pm$ 0.05\\
$T_{\rm eff}$ [K] & 4650 $\pm$ 80 & 6700 $\pm$ 200\\
$\log$(g) (c.g.s.) & 2.70 $\pm$ 0.15 & 4.2 $\pm$ 0.1\\
\hline
 & orbital parameters & \\
 \hline
 $P_{\rm orbit}$ [days] & $>$ 75 & \\
 $a$ [AU] & $>$ 0.5 &\\
\hline
\hline
\end{tabular}
\end{minipage}
\end{table}

From the orbital velocity of the secondary, we computed the period for a
circular orbit, which would be shorter than half the total time span of the TrES
data and we should have seen an occultation in those data. Therefore we
concluded that the orbit is eccentric. The possible annular
eclipse in the SuperWASP data has a length of $\sim$\,7\,days and depth of
$\sim$\,0.03 mag. If confirmed this would indeed imply a large orbital
eccentricity.

From the eclipse times and depth we were able to compute the luminosity and the
radius of the secondary. All values seem to be compatible with an F
main-sequence star. The mass-luminosity relation for main-sequence stars then
leads to a mass estimate of the secondary as well as its surface gravity. These
results are summarized in Table~\ref{allres} and the position of both stars are
indicated in an H-R diagram in Fig.~\ref{fluxosc}.

KIC8410637 is an extremely interesting binary for further follow-up. Longer time
series from \textit{Kepler} will improve dramatically the detection threshold for the
derivation of the properties of the oscillations of the red giant. 

\acknowledgements Funding for this Discovery mission is provided by NASA's Science Mission Directorate. We would like to acknowledge the entire \textit{Kepler} team for their efforts over many years. Without these efforts it would not have been possible to obtain the results presented here. SH, WJC, YPE, IRS and DWK acknowledge support by the UK
Science and Technology Facilities Council. The research leading to these results
has received funding from the European Research Council under the European
Community's Seventh Framework Programme (FP7/2007--2013)/ERC grant agreement
n$^\circ$227224 (PROSPERITY), from the Research Council of K.U.Leuven, from the
Fund for Scientific Research of Flanders (FWO), and from the Belgian Federal
Science Office. DS acknowledges support from the Australian Research Council.

{\it Facilities:} \facility{The Kepler Mission}.


\end{document}